\documentclass[twocolumn]{jpsj2}
\usepackage{graphicx}

\def\runtitle{Current-Induced Entanglement of Nuclear Spins in Quantum Dots}
\def\runauthor{M.\ {\sc Eto}, T.\ {\sc Ashiwa} and M.\ {\sc Murata}}

\title{%
Current-Induced Entanglement of Nuclear Spins in Quantum Dots
}

\author{%
Mikio {\sc Eto},\thanks{E-mail address: eto@rk.phys.keio.ac.jp}
Takashi {\sc Ashiwa} and Mikio {\sc Murata}
}

\inst{%
Faculty of Science and Technology, Keio University, \\
3-14-1 Hiyoshi, Kohoku-ku, Yokohama 223-8522, Japan
}

\recdate{August 5, 2003}

\abst{%
We propose an entanglement mechanism of nuclear spins in quantum dots
driven by the electric current accompanied by the spin flip.
This situation is relevant to a leakage current in spin-blocked regions
where electrons cannot be transported unless their spins are flipped.
The current gradually increases the components of larger total spin of nuclei.
This correlation among the nuclear spins markedly enhances
the spin-flip rate of electrons and hence the leakage current.
The enhancement of the current is observable when the residence time
of electrons in the quantum dots is shorter than the dephasing time
$T_2^*$ of nuclear spins.
}

\kword{%
quantum dot, entanglement, nuclear spins, hyperfine interaction,
spin blockade, dephasing, superradiance, Dicke effect}

\begin{document}
\sloppy
\maketitle

The spin relaxation of electrons in quantum dots is an important issue.
The relaxation time must be sufficiently long for the
implementation of quantum computing devices utilizing electron spins
in quantum dots.\cite{Loss} The time has been examined experimentally by
optical\cite{opt1,opt2} and transport measurements.\cite{Fujisawa,Ono}
Mechanisms of the spin relaxation have been investigated theoretically
by, {\it e.g.}, hyperfine interaction,\cite{Sigurdur,
Sigurdur2,Merkulov,Lyanda-Geller,Khaetskii,Schliemann}
spin-orbit interaction\cite{Khaetskii1,Khaetskii2} and
higher-order tunneling processes (cotunneling).\cite{Averin,Eto}

In this letter, we study the unique properties of nuclear spin states
in quantum dots, which stem from the spin relaxation of electrons
through hyperfine interaction.
In quantum dots, an electron occupying a particular level
couples with numerous nuclear spins simultaneously through the interaction.
Hence a spin flip of the electron results in a correlation among the
nuclear spins. The correlation is an entanglement: we do not know which
of the nuclear spins is flipped.
We show that (i) the electric current accompanied by the spin flip
significantly increases the correlation among nuclear spins, and
(ii) the correlated state of nuclear spins markedly
enhances the spin-flip rate of electrons.
The direct dipole-dipole interaction between the nuclear spins
is much smaller than the hyperfine interaction\cite{Merkulov}
and is thus disregarded. Its influence is discussed later.

This situation is relevant to a leakage current in the spin-blocked region
where electrons cannot be transported unless the spins are flipped. In weakly
coupled double quantum dots, Ono {\it et al}.\ have observed a current
suppression when two electrons occupy the lowest energy level in
each dot with parallel spins;\cite{Ono} an electron tunneling from one dot
to the other is forbidden by the Pauli exclusion principle (see Fig.\ 4).
In this region, we predict that a leakage current
accompanied by the spin flip in one of the dots is significantly enhanced by
the nuclear spin entanglement.
If our theory is verified, it would be the first observation of entanglement
in solid-state devices.

Our mechanism is analogous to the current-induced dynamic nuclear
polarization (DNP) in quantum Hall systems.\cite{DNP1,DNP2,DNP3,Machida}
The DNP is created by electron scattering between spin-polarized edge
states, accompanied by the spin flip. The DNP influences the electronic
state and, in consequence, leads to a hysteresis in the longitudinal
resistance.\cite{Kronmuller,Smet} Note that nuclear spins are entangled but
not polarized in a specific direction in our case.

We examine a simple model for the spin-blocked quantum dots and
discuss its relevance to a realistic situation later.
Our model is as follows.
A quantum dot is connected to external leads, $L$, $R$, through tunnel
barriers.
The Coulomb blockade restricts the number of electrons in the dot to be
$N_{\rm el}$ or $N_{\rm el}+1$. $N_{\rm el}$ electrons form a background
of a spin singlet.
From lead $L$ on the source side, an extra electron tunnels into the dot
and occupies a single level with an envelope wavefunction
$\psi({\mib r})$. The spin of the electron is either $| \uparrow \rangle$
($S_z=1/2$) or $| \downarrow \rangle$ ($S_z=-1/2$) with equal probability
(we assume an easy-axis of electron spins in the $z$ direction, {\it e.g}.,
in a magnetic field). The electron stays for a long time by the spin
blockade, interacting with $N$ nuclear spins,
${\mib I}_k$ (located at ${\mib r}_k$; $k=1,\cdots,N$),
through the hyperfine contact interaction.
After the spin is flipped, the electron tunnels out to lead $R$ on the
drain side, and the next electron is immediately injected from lead $L$.

The hyperfine interaction is expressed as
\begin{equation}
H_{\rm hf} =  2 \sum_{k=1}^N \alpha_k {\mib S} \cdot {\mib I}_k,
\label{eq:Hamil0}
\end{equation}
where $\alpha_k \propto |\psi({\mib r}_k)|^2$.
$N \sim 10^5$ in GaAs quantum dots.
We consider nuclear spins of $1/2$ for simplicity. We assume that
the spin-flip rate from $| \downarrow \rangle$ to $| \uparrow \rangle$
($| \uparrow \rangle$ to $| \downarrow \rangle$) is given by the perturbation
of $H_{\rm hf}$,
\begin{equation}
\Gamma=A |\langle \Psi_{\rm fin}; \uparrow (\downarrow)
|H_{\rm hf}|  \Psi_{\rm init}; \downarrow (\uparrow) \rangle |^2,
\label{eq:Gamma}
\end{equation}
where $\Psi_{\rm init}$ and $\Psi_{\rm fin}$ are the initial and
final states of nuclear spins, respectively.\cite{com0}
We first study the case of homogeneous coupling constants, $\alpha_k = \alpha$.
This is a good approximation for the majority of nuclear
spins since the distance between neighboring nuclei is much smaller than
the size of the quantum dot.
Then the Hamiltonian (\ref{eq:Hamil0}) is reduced to
\begin{equation}
H_{\rm hf} =  2 \alpha {\mib S} \cdot {\mib I},
\label{eq:Hamil}
\end{equation}
where ${\mib I}=\sum_{k=1}^N {\mib I}_k$ is the total spin of nuclei.
A generic case with inhomogeneous hyperfine coupling is examined later.

The Hamiltonian (\ref{eq:Hamil}) [and also (\ref{eq:Hamil0})]
indicates that $N$ nuclear spins interact
with a common ``field'' of an electron spin although there is no direct
interaction between them (dipole-dipole interaction between nuclear spins
is weak and neglected). This field results in an entanglement among
nuclear spins. To illustrate this, let us consider the simplest case of $N=2$
and begin with a polarized state of nuclear spins
$| I_{z}=1/2\rangle_1 | I_{z}=1/2 \rangle_2$. An electron with spin
$| \downarrow \rangle$ tunnels into the dot and is spin-flipped.
Then the state of nuclear spins becomes
$(| -1/2 \rangle_1 | 1/2 \rangle_2 + | 1/2 \rangle_1 | -1/2\rangle_2 )
/\sqrt{2} =| J=1,M=0 \rangle$, where $J$ and $M$ are the total spin and its
$z$ component, respectively. This is an entangled state; we do not know
which of the nuclear spins is flipped.
After the electron tunnels out of the dot, the next electron is injected with
$| \downarrow \rangle$ (or $| \uparrow \rangle$) and interacts with this
state of nuclear spins.
The spin-flip rate for the second electron is given by $2A\alpha^2$.
This value is twice the rate in the case of non-entangled states
of nuclear spins,
$|1/2\rangle_1 | -1/2 \rangle_2$ or $|-1/2\rangle_1 | 1/2 \rangle_2$.
In general, the capability of state $|J,M \rangle$ to
flip an electron spin is $A\alpha^2 (J \pm M)(J \mp M +1)$.

Now we formulate the case of $N$ nuclear spins.
We assume that the initial state of the nuclear spins is
\begin{eqnarray}
\Psi^{(0)} & = & \sum_{I_{z,1}=\pm 1/2} \cdots \sum_{I_{z,N}=\pm 1/2}
C(I_{z,1},\cdots,I_{z,N})
\nonumber \\
& & \times |I_{z,1} \rangle_1 |I_{z,2} \rangle_2 \cdots |I_{z,N} \rangle_N,
\label{eq:Psi00}
\end{eqnarray}
where $\{ C(I_{z,1},\cdots,I_{z,N}) \}$ are randomly
distributed.\cite{com1}
By the transformation of the basis set into the eigenstates of the total
spin $J$ and its $z$ component $M$ of all the nuclei
($J \le N/2$, $-J \le M \le J$), eq.\ (\ref{eq:Psi00}) is rewritten as
\begin{equation}
\Psi^{(0)}=\sum_{J,M,\lambda} C_{J,M,\lambda}^{(0)} | J,M,\lambda \rangle,
\label{eq:Psi0}
\end{equation}
with random coefficients $\{ C_{J,M,\lambda}^{(0)} \}$
($\sum_{J,M,\lambda} |C_{J,M,\lambda}^{(0)}|^2=1$). Index
$\lambda$ distinguishes states with the same $J$ and $M$.
The number of such states is
\begin{equation}
K(N,J)=(2J+1)\frac{N!}{(\frac{N+2J+2}{2})! (\frac{N-2J}{2})!}.
\end{equation}

The first electron with $| \downarrow \rangle$ (or $| \uparrow \rangle$)
enters the dot and interacts with $\Psi^{(0)}$ of nuclear spins. The
spin-flip rate is $\Gamma^{(0)}=A \alpha^2 F^{(0)}$, where
\begin{equation}
F^{(0)}=\sum_{J,M,\lambda} |C_{J,M,\lambda}^{(0)}|^2(J \pm M)(J \mp M+1).
\end{equation}
When $N \gg 1$, $|C_{J,M,\lambda}^{(0)}|^2$ can be replaced by
$1/2^N$ (law of large number; fluctuation of $1/\sqrt{2^N}$ is neglected).
Then $F^{(0)}=(1/2^N)\sum_{J,M} K(N,J) (J \pm M)(J \mp M+1) \approx N/2$.
Note that $\Gamma^{(0)}=(N/2) A \alpha^2$ is identical to the spin-flip rate
which would be evaluated under the assumption that one of the nuclear spins
is flipped with the electron spin.\cite{com3}

\begin{figure}[hbt]
\begin{center}
\includegraphics[width=7.5cm]{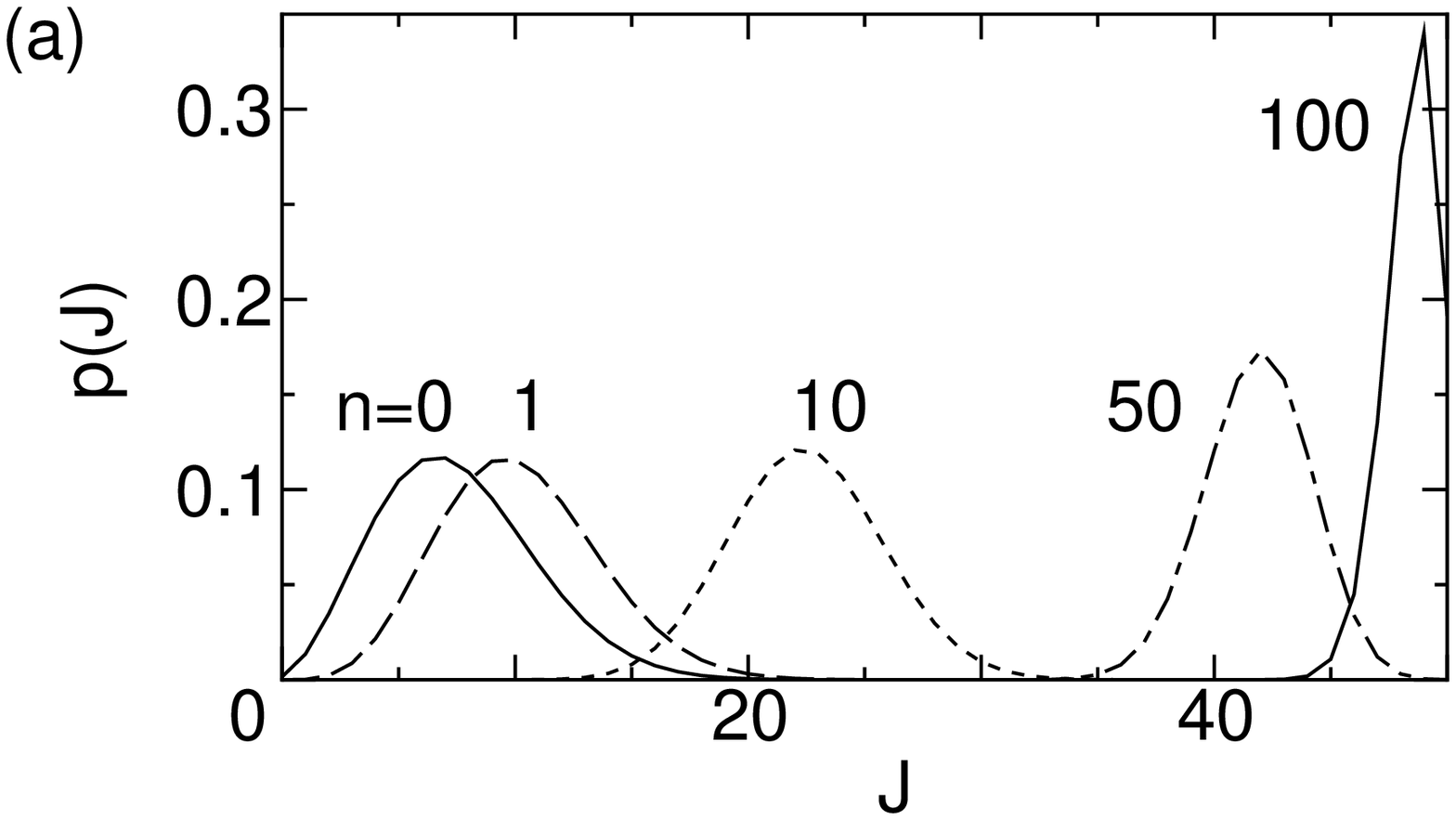} \\
\includegraphics[width=7.5cm]{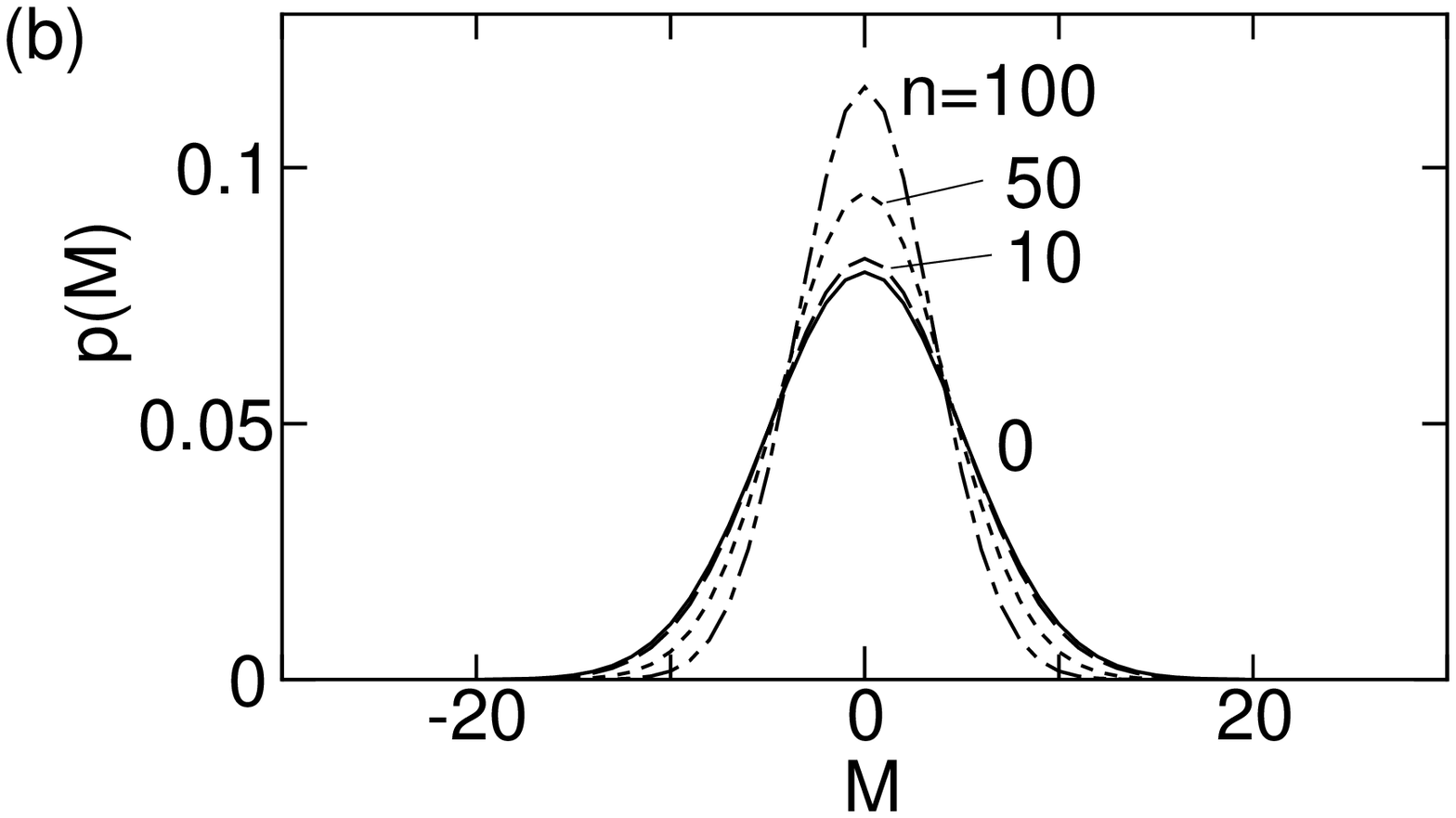}
\end{center}
\caption{Distribution of (a) total spin $J$ and (b) its $z$
component in the state of nuclear spins.
(a) $p(J)=\sum_{M,\lambda} |C_{J,M,\lambda}^{(n)}|^2$, and
(b) $p(M)=\sum_{J,\lambda} |C_{J,M,\lambda}^{(n)}|^2$, where
$n$ is the number of transported electrons accompanied by the spin flip.
The number of nuclear spins is $N=100$. In eq.\ (\ref{eq:Psin}),
we take the geometrical mean between upper and lower signs.}
\end{figure}

After the spin flip, the electron tunnels off the dot.
The state of nuclear spins becomes
\begin{eqnarray}
\Psi^{(1)} = \frac{1}{\sqrt{F^{(0)}}}\sum_{J,M,\lambda} C_{J,M,\lambda}^{(0)}
\sqrt{(J \pm M)(J \mp M+1)}
\nonumber \\
\times | J,M \mp 1,\lambda \rangle.
\end{eqnarray}
$\Psi^{(1)}$ includes more components of larger $J$ and smaller $|M|$.
The correlation among the nuclear spins increases
each time an electron is injected, spin-flipped, and ejected out of the dot.
After $n$ such events, the state of nuclear spins is
\begin{eqnarray}
\Psi^{(n)} & = & \sum_{J,M,\lambda} C_{J,M,\lambda}^{(n)}
                | J,M,\lambda \rangle, \\
\label{eq:Psin0}
C_{J,M \mp 1,\lambda}^{(n)} & = &
\sqrt{\frac{(J \pm M)(J \mp M+1)}{F^{(n-1)}}} C_{J,M,\lambda}^{(n-1)},
\label{eq:Psin}
\\
F^{(n)} & = &
\sum_{J,M,\lambda} |C_{J,M,\lambda}^{(n)}|^2 (J \pm M)(J \mp M+1).
\label{eq:Psin2}
\end{eqnarray}
$F^{(n)}$ is expressed as $F^{(n)} = f^{(n)} / f^{(n-1)}$ with
\begin{equation}
f^{(n)} =  \frac{1}{2^N}\sum_{J,M} K(N,J)
\left[ (J \pm M)(J \mp M+1) \right]^n.
\label{eq:Fn}
\end{equation}
Figure 1 shows the distribution of (a) the total spin $J$ and 
(b) its $z$ component $M$. With increasing $n$, the weight of
larger $J$ and smaller $|M|$ increases. This means that
the total nuclear spins are developed in the plane perpendicular to
the easy-axis of electron spins.

The correlation among the nuclear spins enhances the spin-flip rate.
The rate for the $(n+1)$th electron is given by
$\Gamma^{(n)}=A\alpha^2 F^{(n)}$. Using eq.\ (\ref{eq:Fn}), we find that
$F^{(n)} \approx (N/2)n$ for $1 \ll n \ll N/2$, and
$F^{(n)} \approx (N/2)^2$ for $N/2 \ll n$. Therefore the spin-flip
rate increases with $n$ linearly
($\Gamma^{(n)} \approx n \Gamma^{(0)}$) and finally saturates
($\Gamma^{(n)} \approx (N/2) \Gamma^{(0)}$), as shown in Fig.\ 2(a).

\begin{figure}[hbt]
\begin{center}
\includegraphics[width=7cm]{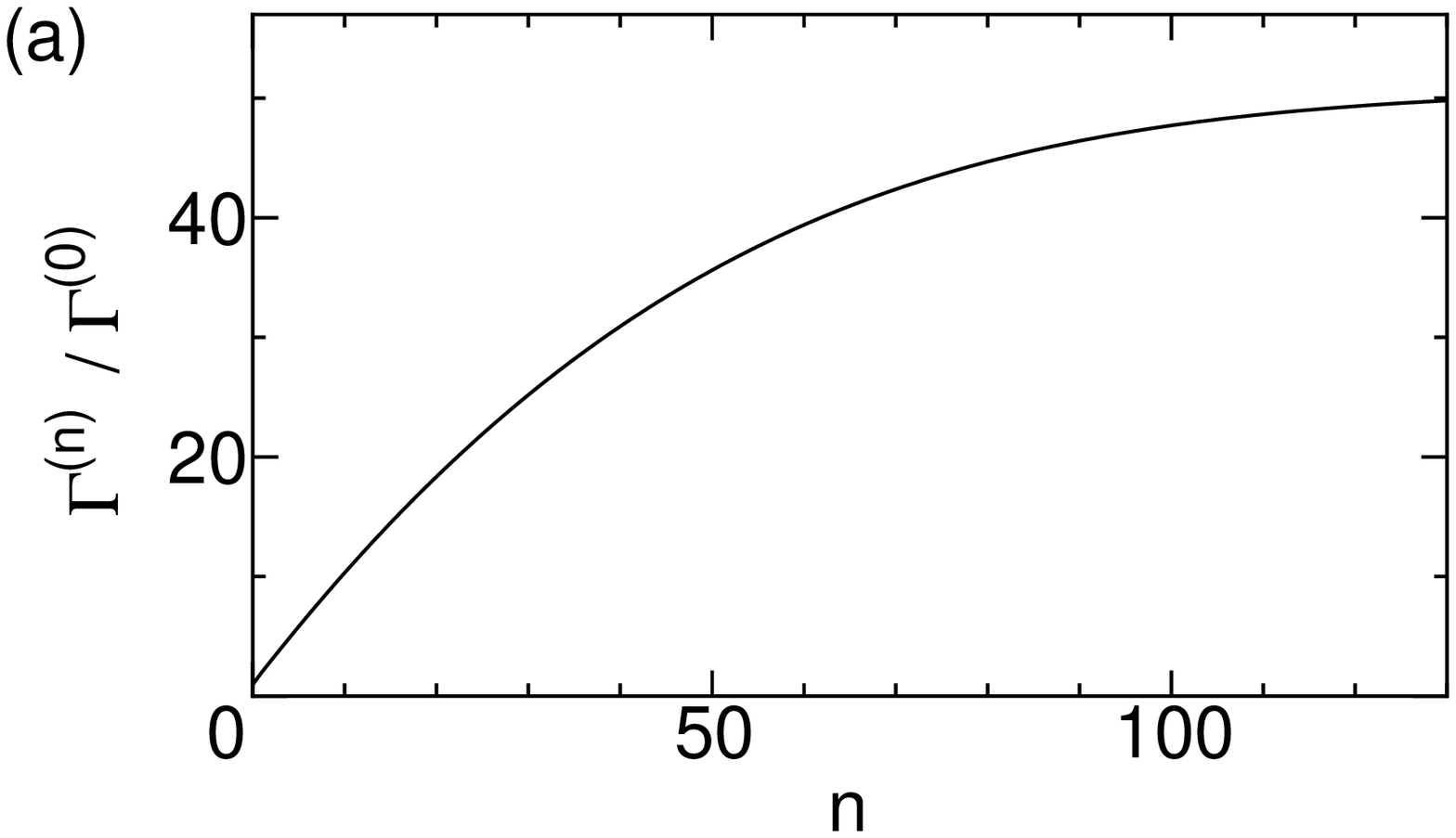} \\
\includegraphics[width=7cm]{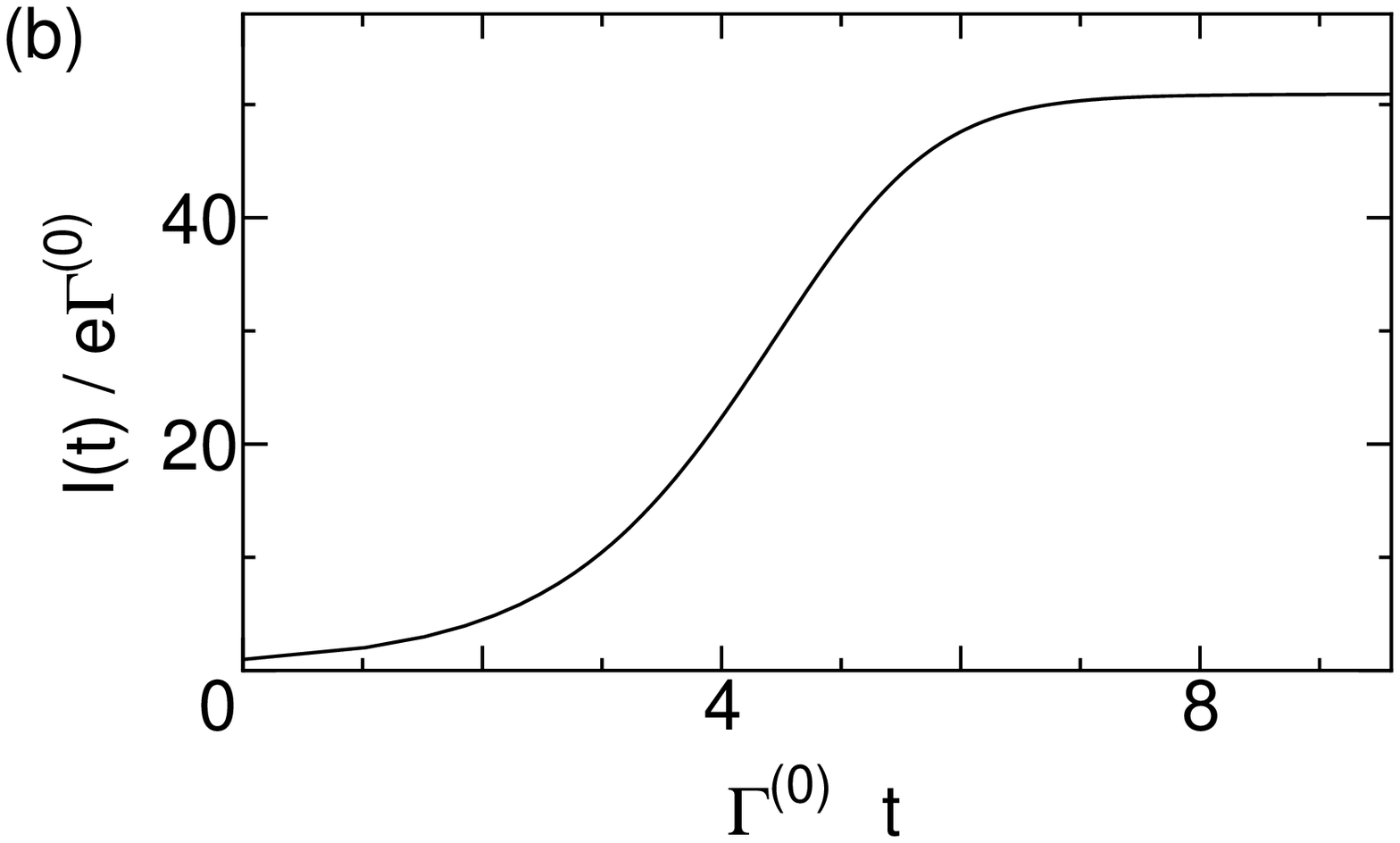}
\end{center}
\caption{(a) Spin-flip rate $\Gamma^{(n)}$
as a function of $n$ (number of transported electrons accompanied by
spin flip) and (b) electric current $I(t)$ as a function of time $t$,
with homogeneous hyperfine coupling.
The number of nuclear spins is $N=100$.
We take the geometrical mean between upper and lower signs in
eq.\ (\ref{eq:Fn}).}
\end{figure}

We evaluate the electric current $I(t)$.
The time interval between $n$th and $(n+1)$th electron transports is
$\Delta t^{(n)} \approx 1/\Gamma^{(n)}$. The current as a function of $t$,
$I(t)=e\Gamma^{(n)}$ at $t=\sum_{j=0}^{n} \Delta t^{(j)}$,
is shown in Fig.\ 2(b). The approximate form of
$\Gamma^{(n)}$ yields an asymptotic form of the current,
\begin{equation}
\left\{
\begin{array}{llll}
I(t) & \approx & e\Gamma^{(0)} e^{\Gamma^{(0)} t}  & \hspace{.5cm}
(t \ll t_{\rm sat}) \\
I(t) & \approx & e(N/2)\Gamma^{(0)} & \hspace{.5cm} (t_{\rm sat} \ll t).
\end{array}
\right.
\label{eq:current}
\end{equation}
The current grows significantly with time and finally saturates.
The saturation time is given by
$t_{\rm sat}=\ln(N/2)/\Gamma^{(0)}$, where $\Gamma^{(0)}$ is the spin-flip
rate of an electron with the uncorrelated state of nuclear spins.

Up to now, we have examined the case of
homogeneous hyperfine coupling in the quantum dot,
$\alpha_k=\alpha$. Generally, the coupling constant
$\alpha_k$ depends on the position of the nucleus due to the spatial
variation of $\psi({\mib r})$. With this inhomogeneous coupling, (i)
the total spin $J$ is no longer a good quantum number, and
(ii) the correlated state of nuclear spins is dephased.
During the stay of an electron in the dot
[$\tau \sim 1/\Gamma^{(0)}=2/(A\alpha^2 N)$],
each term in eq.\ (\ref{eq:Psi00}) gains a phase factor
$e^{i\omega \tau}$ with $\omega=\pm \sum_k \alpha_k I_{z,k} /\hbar$.
This factor diminishes the correlation if the residence time $\tau$ is
sufficiently long.

To examine these effects,
we perform numerical calculations in the case of
$\alpha_k \propto \exp{-\left[ (k-1) / x_0 \right]^2}$ ($k=1,2,\cdots,N$;
$\frac{1}{N}\sum_k\alpha_k=\alpha$).
The number of nuclear spins is $N=14$.
Electrons with spin $| \uparrow \rangle$ and $| \downarrow \rangle$ are
injected into the dot alternately.
For the initial state of nuclear spins, $\Psi^{(0)}$,
the random average is taken over the coefficients
$\{ C(I_{z,1},\cdots,I_{z,N}) \}$ in eq.\ (\ref{eq:Psi00}).

Figure 3 presents the spin-flip rate $\Gamma^{(n)}$ with
$x_0=\infty$ (homogeneous coupling; broken line) and with $x_0=10$ and
$\alpha\tau/\hbar=0$, 10, 20, and 40 (solid lines).
When the dephasing effect is negligible ($\tau=0$),
the spin-flip rate is significantly enhanced by
the entanglement of nuclear spins, even with finite $x_0$. The behavior
of $\Gamma^{(n)}$ is characterized by the effective number of nuclear
spins which couple to the electron spin ($N \rightarrow N_{\rm eff}
\approx x_0$ in the previous discussion).
With an increase in the residence time of electrons in the dot, $\tau$,
the enhancement of
$\Gamma^{(n)}$ becomes less prominent, and is negligible when $\tau$ is
larger than the dephasing time of nuclear spins by the inhomogeneous
hyperfine field.

\begin{figure}[hbt]
\begin{center}
\includegraphics[width=7cm]{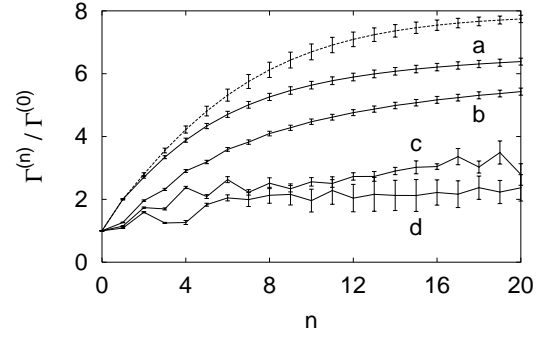}
\end{center}
\caption{Spin-flip rate $\Gamma^{(n)}$ as a function of $n$ (number
of transported electrons accompanied by spin flip),
when the hyperfine coupling constants are
$\alpha_k \propto \exp{\left[ -(k-1) / x_0 \right]}$ ($k=1,2,\cdots,N$;
$\frac{1}{N}\sum_k\alpha_k=\alpha$). $N=14$ with $x_0=10$ (solid lines) or
$\infty$ (broken line). The residence time of the first electron
in the dot, $\tau$, is (a) $\alpha\tau/\hbar = 0$, (b) 10, (c) 20, and (d) 40.
$\Gamma^{(n)}$ is averaged over the random distribution of initial
states. The error bars denote the variation of $\Gamma^{(n)}$.}
\end{figure}

In actual quantum dots, 
there are other origins of the dephasing of nuclear spins, {\it e.g}.,
dipole-dipole interaction. Generally, as long as the dephasing
time $T_2^*$ exceeds $\tau$, our mechanism
appears robust against the dephasing effects because the spin flip of
electrons drives the nuclear spins to be correlated every time.

Our model is relevant to the experimental situation of double quantum dots
shown in Fig.\ 4. A spin blockade takes place when the spin of an incident
electron in dot $L$ is parallel to that of an electron trapped in dot
$R$.\cite{Ono}
We assume a magnetic field in which the Zeeman energy for electrons,
$E_{\rm Z}$, is much larger than $\alpha$ in $H_{\rm hf}$, whereas
the Zeeman energy for nuclear spins
is negligible. The electron-phonon interaction is taken into account for
the energy conservation in the spin flip.\cite{Sigurdur,Sigurdur2}
The Hamiltonian in the dots is
$
H = H_{\rm el}+H_{\rm ph}+H_{\rm T}+H_{\rm hf}+H_{\rm el-ph},
$
where
\begin{eqnarray*}
H_{\rm el} & = & \sum_{j=L,R}
\left[ \sum_{\sigma=\uparrow,\downarrow}(\varepsilon_{j}-E_{\rm Z}S_z)
n_{j,\sigma}+U n_{j,\uparrow}n_{j,\downarrow} \right], \\
H_{\rm ph} & = & \sum_{\mib q}\hbar\omega_{\mib q} b^{\dagger}_{\mib q}b_{\mib q},
\\
H_{\rm T} & = & T\sum_{\sigma=\uparrow,\downarrow}
(a^{\dagger}_{R,\sigma}a_{L,\sigma}+a^{\dagger}_{L,\sigma}a_{R,\sigma}), \\
H_{\rm el-ph} & = & \sum_{\mib q} (\beta_{L,\mib q}n_{L}+\beta_{R,\mib q}n_{R})
(b^{\dagger}_{\mib q}+b_{- \mib q}),
\end{eqnarray*}
where $a^{\dagger}_{j,\sigma}$, $a_{j,\sigma}$
($b^{\dagger}_{\mib q}$, $b_{\mib q}$)
are creation and annihilation operators, respectively,
for an electron in dot $j$
(a phonon), $n_{j,\sigma}=a^{\dagger}_{j,\sigma} a_{j,\sigma}$
and $n_{j}=n_{j,\uparrow}+n_{j,\downarrow}$.\cite{Brandes0}

In the spin blockade of $|L \uparrow \rangle |R \uparrow \rangle$,
an electron tunneling to $|R \downarrow \rangle |R \uparrow \rangle$
is accompanied by the spin flip in one of the dots, {\it e.g}.,
dot $L$, and emission
of a phonon of $\hbar\omega_{\mib q}=\delta E=
\varepsilon_{L}-(\varepsilon_{R}+U)-E_{\rm Z}$. The transition rate is
$\Gamma=A |\langle \Psi_{\rm fin}^{(L)}; \downarrow
|H_{\rm hf}|  \Psi_{\rm init}^{(L)}; \uparrow \rangle |^2$ with
\[ 
A=
\frac{2\pi}{\hbar}
\left( T \frac{\beta_{L,\mib q}-\beta_{R,\mib q}}{E_{\rm Z} \delta E} \right)^2
D_{\rm ph}(\delta E) \left[ N_{\rm ph}(\delta E)+1 \right]
\] 
in the lowest order of $H_{\rm T}+H_{\rm hf}+H_{\rm el-ph}$.
Here, $D_{\rm ph}$ is the density of states for phonons and $N_{\rm ph}$
is the number of phonons.\cite{com2}
This agrees with eq.\ (\ref{eq:Gamma}).

\begin{figure}[hbt]
\begin{center}
\includegraphics[width=6cm]{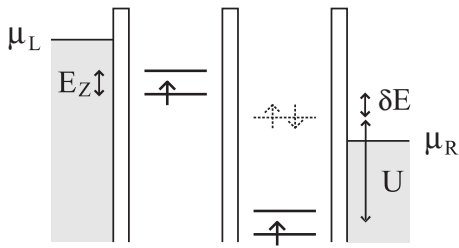}
\end{center}
\caption{An experimental situation of double quantum dots.
A spin blockade takes place when the spin of an incident electron in
dot $L$ is parallel to that of the electron trapped in dot $R$.\cite{Ono}
The leakage current in the spin-blocked region is relevant to our theory.
}
\end{figure}

Besides eq.\ (\ref{eq:Gamma}), we have assumed that the nuclear spin state is
a pure state after an electron tunnels out from dot state
$|R \downarrow \rangle |R \uparrow \rangle$ to lead $R$ and is detected by
current measurement.\cite{com0}
For this assumption to be valid, the following higher-order tunneling processes
(cotunneling)\cite{Averin,Eto,Fujisawa} must be disregarded:
(i) third-order processes in which an electron tunnels out from dot $R$ to lead
$R$, an electron tunnels from dot $L$ to dot $R$, and finally an electron
tunnels into dot $L$ from lead $L$, or (ii) second-order processes in which
an electron tunnels out from dot $L$ ($R$) to lead $L$ ($R$), and another
electron tunnels in from lead $L$ ($R$) to dot $L$ ($R$).
In the presence of such processes and hyperfine couplings,
the nuclear spin state would be entangled with electron states in the leads.
Then our calculations would no longer be justified.

We have also neglected the spin-orbit (SO) interaction, which
may play a role in the spin relaxation of electrons in quantum
dots.\cite{Khaetskii1,Khaetskii2} The coexistence of hyperfine and SO
interactions would complicate the evolution of the nuclear spin state
although the state remains pure (a linear combination of spin-flipped and
-unflipped states) in this case.
Its analysis is beyond the scope of the present study.

In conclusion,
we have proposed a current-induced entanglement of nuclear spins in a
quantum dot. The current accompanied by spin flip in the quantum dot
gradually increases the correlation among nuclear spins when the
residence time of electrons in the dot does not exceed the dephasing time
of nuclear spins $T_2^*$. The correlated state of nuclear spins
significantly enhances the spin-flip rate of electrons.
A relevant situation to this theory is the leakage current in
spin-blocked regions.

The time dependence of the leakage current
$I(t)$ shown in Fig.\ 2(b) may be observed in the experiment
using a sequence
of NMR pulses. The nuclear spins are randomized by each pulse.
By tuning the interval of the pulses, $\Delta t$, $I(\Delta t)$ can be
measured.
Quantitatively, the dephasing time of nuclear spins is estimated to be
$T_2^* \sim 1$ ns in a case of GaAs quantum dots.\cite{Merkulov}
The saturation time of $I(t)$ is of the order of the residence time of
electrons in quantum dots [$t_{\rm sat}=\ln(N/2) \tau$ in eq.\
(\ref{eq:current})], which should be shorter than $T_2^*$.

Finally, we comment on an analogy between our mechanism and the Dicke effect of
superradiance.\cite{Dicke,Brandes} In the spontaneous emission of
photons from $N$ atoms with two levels (pseudo-spin $S_z=\pm 1/2$),
all the atoms could interact with a common electromagnetic field.
The emission of photons is significantly enhanced if $N$ atoms are
excited initially. This is due to the formation of the pseudo-spin state
$|J,M \rangle$ with $J=N/2$. Starting from $|J,J \rangle$,
the state of $N$ atoms changes like a cascade, $|J,J \rangle$,
$|J,J-1 \rangle, \cdots$, emitting photons rapidly.
A similar effect has been proposed for the emission of phonons from $N$
equivalent quantum dots.\cite{Brandes,Brandes2}
The atoms (quantum dots) correspond to the nuclear spins in our model, and
the emission of photons (phonons) to the spin flip of electrons.
A major difference is the initialization.
$N$ excited states must be prepared by pumping in the Dicke effect,
whereas such initialization is not necessary in our mechanism.

The authors are indebted to L.\ P.\ Kouwenhoven for the suggestion of
the experiment using NMR pulses. They gratefully acknowledge discussions with
K.\ Kawamura, R.\ Fukuda, S.\ Komiyama, T.\ Inoshita, A.\ Shimizu,
K.\ Ono, G.\ E.\ W.\ Bauer, and S.\ I.\ Erlingsson.
This work was partially supported by a Grant-in-Aid for
Scientific Research in Priority Areas ``Semiconductor Nanospintronics''
(No.\ 14076216) of the Ministry of Education, Culture, Sports, Science
and Technology, Japan.


\begin{thebibliography}{99}
\bibitem{com0}
We assume that the final state of nuclear spins is a pure
state, not a mixed state with electron states $| \uparrow \rangle$ or
$| \downarrow \rangle$, after the electron tunnels out of the dot.
We observe a spin flip of electron by the leakage current in a spin-blocked
situation, whereas we cannot observe which of the nuclear spins was flipped.
In a spin-flip process, a phonon may be emitted. However, if the wavelength
of the phonon is much longer than the nucleus-nucleus distance,
we do not obtain such information on the nuclear spins.
\bibitem{com3}
Hence the entanglement of nuclear spins cannot be observed
by optical experiments in which the spin relaxation of single excitations
is measured.\cite{opt1,opt2}
\bibitem{com2}
For the transition rate from $|L \downarrow \rangle |R \downarrow \rangle$
to $|R \uparrow \rangle |R \downarrow \rangle$, $\delta E$ is replaced
by $\delta E'=\delta E+2E_{\rm Z}$. Note that
the spin of an incident electron in dot $L$ is parallel or anti-parallel
to the spin of an electron trapped in dot $R$, with equal probability.
In the latter case,
the spin blockade does not occur; the electron tunnels in the second dot
and finally out to lead $R$ immediately. The current is the sum of the
currents with and without spin flip.
\end{thebibliography}
\end{document}